\documentclass{mem}
\usepackage{natbib}
\usepackage{txfonts}
\usepackage{balance}
\usepackage{graphicx}
\usepackage[a4paper,breaklinks,dvipdfm]{hyperref}
\idline{75}{282}

\begin{document}
\def\teff{$T\rm_{eff }$}
\def\kms{$\mathrm {km s}^{-1}$}

\title{
Investigating the magnetism of Brown Dwarfs
}

   \subtitle{}

\author{
O. \,Kuzmychov\inst{1} 
\and S.V. \,Berdyugina\inst{1,2}
\and D. \,Harrington\inst{1,2}
\and J. \,Kuhn\inst{2}
          }

  \offprints{O. Kuzmychov}

\institute{
Kiepenheuer-Institut f\"ur Sonnenphysik, 
Sch\"oneckstr. 6, 
79104 Freiburg, 
Germany 
\and
Institute for Astronomy, 2680 Woodlawn Drive, Honolulu HI 96822-1839, USA\\
\email{oleksii.kuzmychov@kis.uni-freiburg.de}
}

\authorrunning{Kuzmychov}

\titlerunning{Investigating the magnetism of BDs}

\abstract{
We model the spectra of two brown dwarfs observed with the low resolution spectropolarimeter LRIS ($R \sim 3000$, Keck observatory) during several rotational phases in order to infer their magnetic properties. The spectra modeled include the intensity signal (Stokes $I/I_c$) as well as the polarimetric signals (Stokes $Q/I_c$, $U/I_c$ and $V/I_c$), all coming from the 0-0 vibrational band of the CrH molecule at $\approx 8610$ \AA. \\
In order to model the Stokes profiles, we solve a set of the radiative transfer equations for the CrH transitions in the presence of an external magnetic field.\\
We present here the upper limits for the magnetic field strengths for the objects observed, based on the modeling of the intensity signal $I/I_c$ and the signal-to-noise information only. The proper modeling of the polarimetric signals, that requires more careful data reduction, is underway. Nevertheless, our preliminary results show a hint for kG magnetic fields for both brown dwarfs, that is in a good agreement with the result obtained from the simultaneous radio, H$\alpha$ and X-Ray observations of one another radio pulsating brown dwarf. 

}

\maketitle{}

\section{Introduction}
The evidences for the stellar magnetic activity have been found across the entire HR diagram, from cool dwarfs to rapidly rotating giants. There are handful of techniques that allow to explore the stellar magnetism \citep[see][]{berdyugina2005}.\\
\indent In cool stars, the Zeeman broadening measurements are a wide used tool for inferring the information on the magnetic field strength \citep{robinson1980, saar1988, valenti2001}. Because the atomic lines become weak for \teff\, $\lesssim 3000$ K, this technique does not work well for cool M dwarfs and brown dwarfs. However, some molecular lines can be used for Zeeman broadening measurements in the stellar atmospheres cooler than \teff\, $=3000$~K. Thus, \citet{valenti2001_feh} and \citet{berdyugina2001} suggested the FeH molecule for this purpose. The capabilities of the FeH for measuring of the stellar magnetic fields was investigated by \citet{berdyugina2002} and \citet{berdyugina2003}. An overview over the molecules, that can be used for magnetic field measurements in cool stars, can be found in  \citet{bernath2009}. \\
\indent The magnetism of brown dwarfs is not well studied yet. However, a small number of brown dwarfs that exhibit transient but periodic radio flares have been observed simultaneously in radio, H$\alpha$ and X-Rays \citep[see][]{berger2005}. These radio flares are believed to be driven by the electron cyclotron maser mechanism, for which a kG magnetic field is required \citep{hallinan2006, hallinan2007}.\\
\indent In this work, we are going to explore the magnetism of two radio pulsating brown dwarfs with help of a spectropolarimetric technique developed for the CrH molecule by \citet{kuzmychov2012, kuzmychov2013}. This technique makes use of the fact that in the presence of a kG magnetic field the CrH lines produce a measurable broad-band polarimetric signal.\\
\indent The outline of the paper is as follows. In Sec.~\ref{sec: observations}, we provide information on the data collected and the data reduction steps undertaken. Section~\ref{sec: modeling} describes briefly how we calculate a synthetic CrH spectrum we are going to model to the data. Then, in Sec.~\ref{sec: results}, we present our very preliminary results on the magnetic field strength and field distribution for both brown dwarfs observed. Finally, we will draw conclusions from this work in Sec.~\ref{sec: conclusions}.

\section{Observations and data reduction}\label{sec: observations}
We conducted the full Stokes measurements of two radio pulsating brown dwarfs -- M$8.5$ object \object{2MASS J18353790+3259545} (hereafter \object{2M 1835+32}) and L$3.5$ object \object{2MASS J00361617+1821104} (hereafter \object{2M 0036+18}) -- at different rotational phases during two nights in August 2012. The observations were done with the low resolution dual-beam spectropolarimeter LRIS ($R \sim 3000$) at Keck observatory. With $\approx 10$ min exposure time, we achieved the signal-to-noise ratio of $\sim 300$ and $\sim 100$ for M$8.5$ and L$3.5$ object, respectively. \\
\indent In order to complete one full Stokes measurement ($Q/I_c$, $U/I_c$ and $V/I_c$), a sequence of six exposures is done. Due to the beamsplitter, one exposure gives two 2D spectra that have orthogonal polarization states. In order to minimize the instrumental effects, a second exposure for each of three Stokes parameters is done by rotating the retarder by $45^\circ$, so that now the polarization states of the beams are exchanged. One Stokes parameter is then calculated by averaging over a set of two subsequent exposures as follows \citep[see e.g.][]{harrington2008}:
\begin{equation}
q=\frac{1}{2}\left(\frac{i^1_{\shortparallel,0^\circ}-i^1_{\perp,0^\circ}}{i^1_{\shortparallel,0^\circ}+i^1_{\perp,0^\circ}} - \frac{i^2_{\shortparallel,45^\circ}-i^2_{\perp,45^\circ}}{i^2_{\shortparallel,45^\circ}+i^2_{\perp,45^\circ}}\right),
\label{eq: stokesq}
\end{equation}
\begin{equation}
u=\frac{1}{2}\left(\frac{i^3_{\shortparallel,22.5^\circ}-i^3_{\perp,22.5^\circ}}{i^3_{\shortparallel,22.5^\circ}+i^3_{\perp,22.5^\circ}} - \frac{i^4_{\shortparallel,67.5^\circ}-i^4_{\perp,67.5^\circ}}{i^4_{\shortparallel,67.5^\circ}+i^4_{\perp,67.5^\circ}}\right),
\end{equation}
\begin{equation}
v=\frac{1}{2}\left(\frac{i^5_{\shortparallel,0^\circ}-i^5_{\perp,0^\circ}}{i^5_{\shortparallel,0^\circ}+i^5_{\perp,0^\circ}} - \frac{i^6_{\shortparallel,45^\circ}-i^6_{\perp,45^\circ}}{i^6_{\shortparallel,45^\circ}+i^6_{\perp,45^\circ}}\right).
\label{eq: stokesv}
\end{equation}
where $i$ is the flux recorded by the CCD, and its upper index denotes the exposure's number. Thus, for measuring the Stokes $Q/I_c$ two exposures are done: one with the half-wave plate at $0^\circ$, that gives the both intensities $i^1_{\shortparallel,0^\circ}$ and  $i^1_{\perp,0^\circ}$, and another one with the half-wave plate at $45^\circ$, that gives $i^2_{\shortparallel,45^\circ}$ and $i^2_{\perp,45^\circ}$. For measuring the Stokes $U/I_c$, the orientations of the half-wave plate are $22.5^\circ$ and $67.5^\circ$. When measuring the Stokes $V/I_c$, a quater-wave plate at $0^\circ$ and $45^\circ$ is used. The Stokes $I/I_c$ can be obtained from each of the Stokes measurements (\ref{eq: stokesq}) - (\ref{eq: stokesv}) in the following way:
\begin{equation}
i=\frac{1}{2}\left(\frac{i^n_{\shortparallel,\alpha}+i^n_{\perp,\alpha}}{2} + \frac{i^{n+1}_{\shortparallel,\alpha+45^\circ}+i^{n+1}_{\perp,\alpha+45^\circ}}{2}\right),
\label{eq: stokesi}
\end{equation}
where $n=1,3,5$ is the exposure's number, and $\alpha=0^\circ$ for both $Q/I_c$ and $V/I_c$ and $\alpha = 22.5^\circ$ for $U/I_c$.

\indent Because there is no public data reduction pipeline for LRIS, we developed our own routines that extract the spectra from the CCD images. Before extracting the spectra, the CCD images were corrected for distortions that arise when the light passes through the optics before hitting a CCD. Furthermore, we subtracted the flat field and the sky background from the data. The wavelength calibration was done with an Arc lamp spectrum recorded during the observing run. In order to account for the instrumental polarization, unpolarized and polarized standard stars were observed to compute zero-point and orientation information.  As an additional efficiency and absolute calibration test, observations using of the daytime sky were carried out following \citet{harrington2011}. \\
\indent One another data reduction step we are still working on is removal of the cosmic ray hits. In order to keep the accuracy of the polarimetric signal observed near the level given by the poisson statistics and the instrument performance, the proper cleaning of the data from the cosmic particle hits is necessarily.

\section{Modeling}\label{sec: modeling}
\begin{figure}[]
\resizebox{\hsize}{!}{\includegraphics[clip=true]{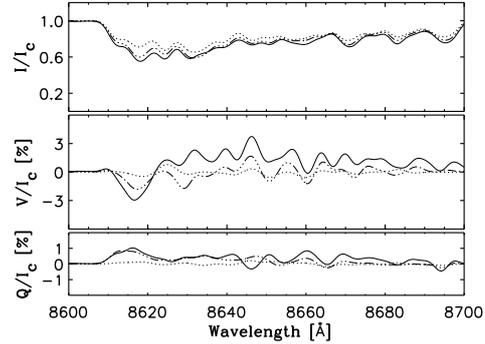}}
\caption{Stokes profiles simulated for the \mbox{0-0} band of the CrH molecule for 1 kG (dotted line), 3 kG (dotted-dashed line) and 6 kG (solid line) magnetic field strengths. The filling factor of the magnetic elements is not taken into account (assumed to be $1$). The orientation of the magnetic field is $\gamma = 45^\circ$ and $\chi = 0^\circ$, where $\gamma$ is the inclination and $\chi$ is the azimuth angle of the field with respect to the line of sight. We use here a Phoenix atmospheric model calculated for $T_\textrm{eff}=2500$~K, $\log g=4.5$ and solar abundances of the chemical elements. The instrumental broadening $3.5$ \AA, which corresponds to that of LRIS, and the broadening due to stellar rotation $v\sin i=60$~\kms\, were taken into account.}
\label{fig: 2500K}
\end{figure}

We calculate the four Stokes parameters ($I/I_c$, $Q/I_c$, $U/I_c$ and $V/I_c$) for the \mbox{0-0} band from the $A^6\Sigma - X^6\Sigma$ system of the CrH in the presence of a magnetic field according to \citet{kuzmychov2012, kuzmychov2013}. For doing this, we employ the radiative transfer code STOPRO that calculates the Stokes parameters  of one ore more spectral lines \citep{solanki1987, berdyugina2003}. On input, the code requires the molecular (or atomic) data of the spectral lines to be calculated and an atmospheric model. We use both Phoenix \citep{allard95} and Drift-Phoenix \citep{witte09} atmospheric models, depending on the \teff\, of the object which spectrum we are interested in.\\
\indent Figure \ref{fig: 2500K} shows the Stokes profiles simulated for the \mbox{0-0} band of the CrH molecule at the magnetic field strengths $1-6$~kG. The stronger is the magnetic field, the bigger is the polarimetric signal, reaching a few percent in Stokes $V/I_c$ at $6$~kG. Furthermore, in a kG field regime both $Q/I_c$ and $V/I_c$ show a substantial asymmetry over the wide range of wavelengths, the so-called \emph{broad-band polarization}. This asymmetric shape of the polarimetric signals does not depend on the filling factor, but only on the magnetic field strength (filling factor will merely reduce the strength of the signal, but it will not affect its shape).


\section{Results}\label{sec: results}
\begin{figure}[]
\resizebox{\hsize}{!}{\includegraphics[clip=true]{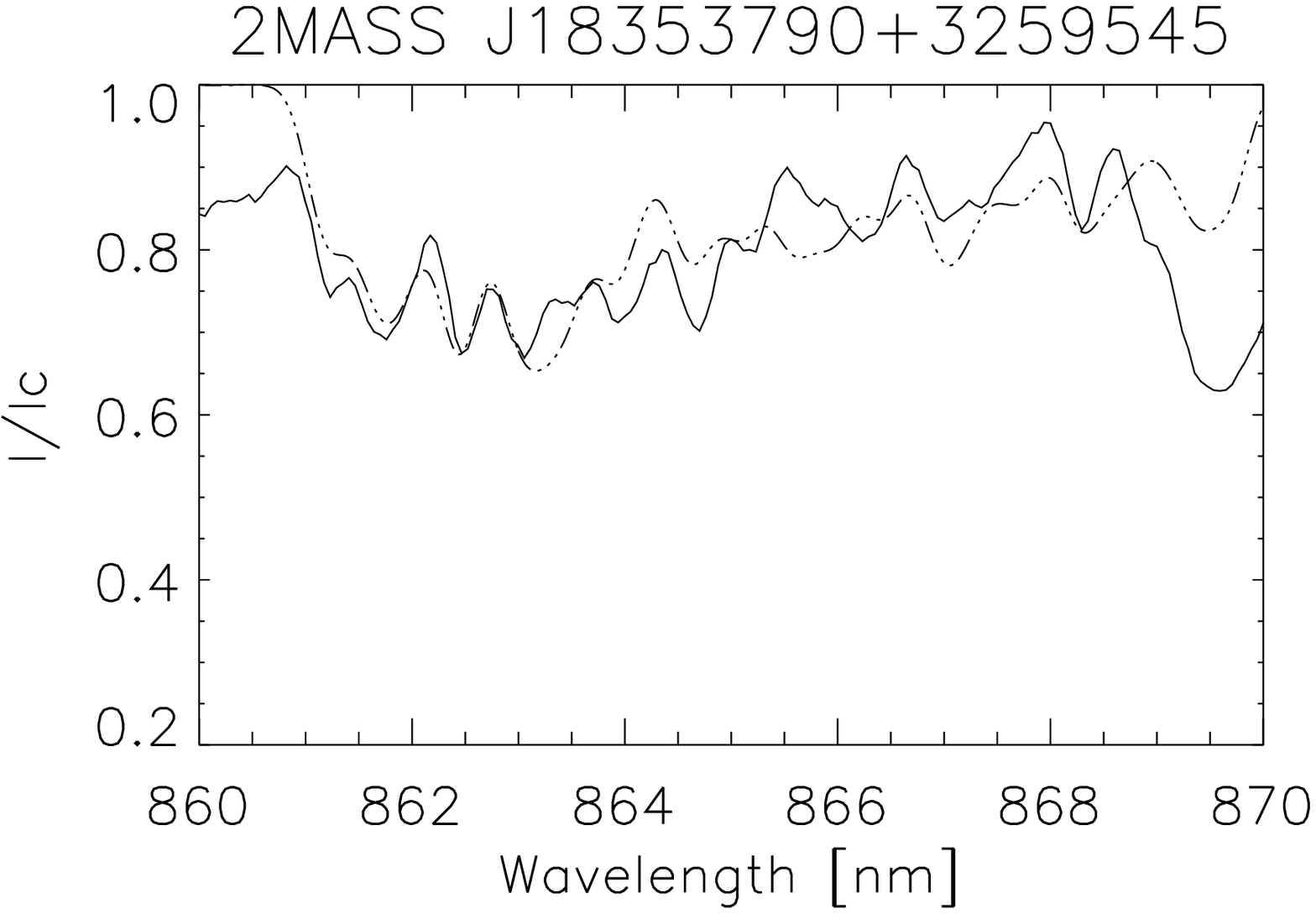}}\\
\resizebox{\hsize}{!}{\includegraphics[clip=true]{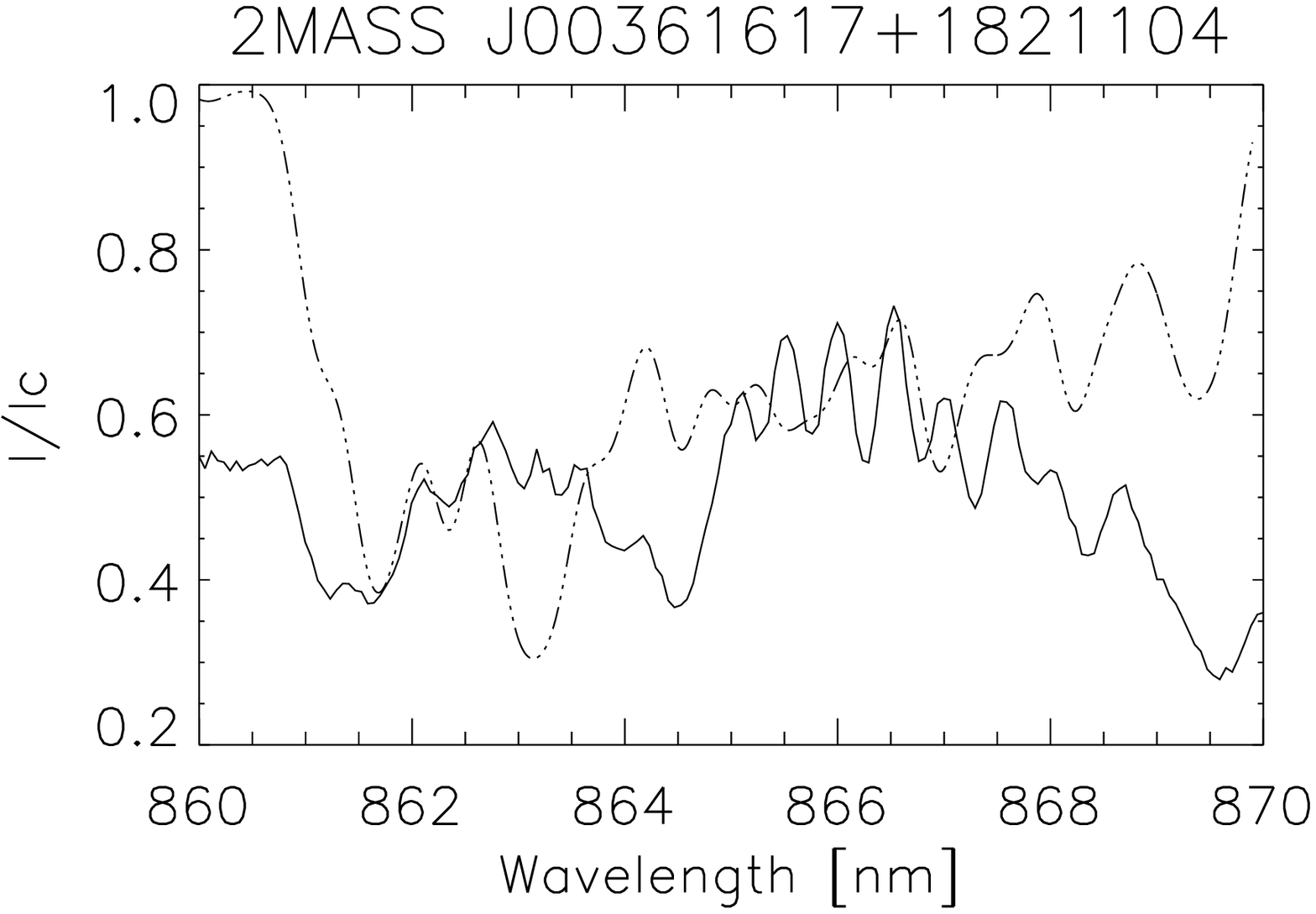}}
\caption{Stokes profiles $I/I_c$ for the \mbox{0-0} band of the CrH (solid line) modeled to the intensity signals observed (dashed-dotted line). For \object{2M 1835+32} object (upper panel), we calculate $I/I_c$ for $1$~kG magnetic field strength using a Drift-Phoenix atmospheric model for \teff\, $=2500$~K. $I/I_c$ for \object{2M 0036+18} object (lower panel) is calculated for $3$~kG magnetic field using a Drift-Phoenix atmospheric model for \teff\, $=2000$~K. Both atmospheric models were obtained for $\log g=5.0$ and solar abundances of the chemical elements. The orientation of the magnetic field is as indicated in the caption to Fig. \ref{fig: 2500K}.  In both cases, we take account of the instrumental broadening of $3.5$~\AA, which corresponds to that of LRIS, and the broadening due to stellar rotation of $v\sin i=60$~\kms.}
\label{fig: 2mass}
\end{figure}

Here we model the synthetic spectra of the \mbox{0-0} band of the CrH in order to match it to the data observed (see Sec. \ref{sec: observations}). Because for the proper modeling of the polarimetric signal a more careful data reduction is required,  the results presented here are based on the modeling the intensity signal and the signal-to-noise information only. \\
\indent Figure~\ref{fig: 2mass} shows the best match of the modeled Stokes $I/I_c$ to the intensity spectra observed. For \object{2M 18353+32} object, the spectrum modeled shows a good correspondence to the observed one (cf. Fig.~\ref{fig: 2mass}, upper panel). The region between roughly $863$ and $867$~nm, where the both curves do not match as good, is most likely contaminated by the TiO lines. One another diatomic molecule, which lines contaminate the CrH spectrum on its red side, is FeH. The head of its \mbox(1,0) vibrational band from the $F^4\Delta - X^4\Delta$ system lies around $869$~nm \citep{carroll1976}.\\
\indent The modeling of the intensity signal for \object{2M 0036+18} object, that is significantly cooler than the other one, is much more challenging (cf. Fig.~\ref{fig: 2mass}, lower panel). Not only the lines of the CrH, but also that of TiO and FeH that blend with the \mbox{0-0} band of the CrH, become very deep and broad. Nevertheless, using only the CrH lines we could simulate a signal, that reproduces at least a couple of features of the spectrum observed. Thus, the band head is clearly identified, even though it appears narrower in the simulated signal than in the observed one. A possible reason for this could be the TiO lines missing in our model. Also, the bumps around $862.5$ and $866.8$~nm definitely belong to the CrH.\\
\indent Based on the comparison of the Stokes $I/I_c$ simulated with the observational data, we can get an idea on the magnetic field strength and its spatial distribution for both objects observed. Furthermore, from the signal-to-noise information of the data, we can deduce the upper limit of the magnetic field strength multiplied with the filling factor. This results are collected in Tab. \ref{tab: result}. 

\begin{table}
\caption{Upper limits for the magnetic field strengths.}
\begin{center}
\begin{tabular}{lll}
Object&Field, kG&ff\\\hline
               	&$1$&$1$\\
 \object{2M 1835+32} (M$8.5$) &$3$&$0.6$\\
			&$6$&$0.3$\\\hline
			&$1$&$1$\\
\object{2M J0036+18} (L$3.5$) &$3$&$0.5$\\
			&$6$&$0.3$
\label{tab: result}
\end{tabular}
\end{center}
\end{table}


\section{Conclusions}\label{sec: conclusions}
From the results obtained in the preceding section, we can draw the following conclusions.\\
\indent The spectropolarimetric technique for the CrH, that was developed by \citet{kuzmychov2012, kuzmychov2013} and that is applied here for the first time to the spectra of brown dwarfs, showed itself in practice. Even though we did not fully exploit the potential of this technique, it yields very promising predictions for the magnetic field strength and its spatial distribution for both objects observed.\\
\indent Our calculations indicate the existence of a kG surface magnetic field for both brown dwarfs considered. This results are in a good agreement with that inferred by \citet{hallinan2006, hallinan2007} from the simultaneous radio, H$\alpha$ and X-Ray observations of one another radio pulsating brown dwarf.\\
\indent After having reduced the data properly, we are going to model the polarimetric signals. This would provide us with the more reliable information on the magnetic field properties of the objects observed. We are also going to include into our calculations the lines of the TiO and FeH, that blend with the \mbox{0-0} band of the CrH.

\begin{acknowledgements}
We are thankful to Ch. Helling for providing us with the Drift-Phoenix atmospheric models for \teff\, $=2000$~K and $2500$~K.
\end{acknowledgements}

\bibliographystyle{aa}

\end{document}